\documentclass[11pt]{amsart}

\usepackage{hyperref}

\usepackage{enumitem}

\usepackage{xfrac}  
\usepackage{mathrsfs}		

\usepackage[round]{natbib}

\newtheorem{thm}{Theorem}[section]

\theoremstyle{definition}

\theoremstyle{remark}

\newtheorem{exa}[thm]{Example}

\newtheorem{exe*}[thm]{Exercise*}
\newtheorem{exe!}[thm]{Exercise(!)}
\numberwithin{equation}{section}
\newcommand{\absco}{{<\kern-0.53em<}}

\newcommand{\dbracc}[1]{[\kern-0.15em[ #1 ]\kern-0.15em]}
\newcommand{\dbraoc}[1]{]\kern-0.15em] #1 ]\kern-0.15em]}
\newcommand{\dbraco}[1]{[\kern-0.15em[ #1 [\kern-0.15em[}
\newcommand{\dbraoo}[1]{]\kern-0.15em] #1 [\kern-0.15em[}

\title{Diversifying an Index}

\author{Johannes Ruf}

\address{Johannes Ruf\\
  Department of Mathematics\\
  London School of Economics and Political Science}

\email{j.ruf@lse.ac.uk}

\date{July 13, 2023}
\begin{document}

\begin{abstract}
In July 2023, Nasdaq announced a `Special Rebalance' of the Nasdaq-100 index to reduce the index weights of its large constituents.  A rebalance as suggested currently by Nasdaq index methodology may have several undesirable effects.  These effects can be avoided by a  different, but simple rebalancing strategy.  Such rebalancing is easily computable and guarantees (a) that the maximum overall index weight does not increase through the rebalancing and (b) that the order of index weights is preserved.
\end{abstract}
\maketitle

\section{Background}
As of July 2023, the top 6 constituents in the Nasdaq-100 index have an aggregate index weight of over 50\%.  Nasdaq\footnote{\url{https://www.nasdaq.com/press-release/the-nasdaq-100-index-special-rebalance-to-be-effective-july-24-2023-2023-07-07}} decided to address this overconcentration in the index by redistributing the index weights, in line with its official index methodology\footnote{\url{https://indexes.nasdaq.com/docs/Methodology_NDX.pdf}}. Although the rebalancing methodology is not explicitly laid out  for the situation of a `Special Rebalance', its standard weighting process indicates the following procedure: The sum of all index weights that are larger than 4.5\% gets reweighted to 40\%. 

Matt Levine in his excellent newsletter\footnote{\url{https://www.bloomberg.com/opinion/articles/2023-07-12/it-s-easy-to-make-oil-companies-esg\#xj4y7vzkg}} described several undesirable effects that Nasdaq's procedure might cause:
\begin{itemize}
	\item[(a)] The ordering of index weights is not preserved. A constituent with a currently higher index weight might have a lower index weight after the rebalancing. 
	\item[(b)] The largest index weight after the rebalancing might be larger than the largest index weight before the rebalancing.
\end{itemize}

\section{Proposed methodology}
Currently, each index weight is a relative market capitalization, i.e., constituent share price times their total shares outstanding divided by the aggregated market capitalization. Let us write $\mu_1, \mu_2, \ldots, \mu_{100}$ to denote these relative market capitalizations. By definition, they are all  nonnegative and sum up to one: $\mu_1 + \mu_2 + \cdots + \mu_{100} = 1$.  Hence, they are valid index weights.  

Let us write $\eta_1, \eta_2, \ldots, \eta_{100}$ for the new market weights. 
Nasdaq's standard reweighting might lead to the situation that
\begin{itemize}
	\item[(a)] $\mu_i < \mu_j$  but $\eta_i > \eta_j$ for some numbers $i$ and $j$ between 1 and 100. 
	\item[(b)] The maximum of all $\eta_i$'s is larger than the maximum of all $\mu_i$'s.
\end{itemize}

For the proposed reweighting, let us consider some number $p$ between 0 and 1.  We then set the new index weights equal to the original ones to the power of p and normalized to one, i.e.,
$$
	\eta_i = \frac{\mu_i^p}{\mu_1^p + \mu_2^p + \cdots + \mu_{100}^p}.
$$ 
For such choice of reweighting one always is guaranteed to preserve ordering and the maximum index weight is not increasing.

One is free to choose  $p$ between 0 and 1.  If $p=1$, one obtains the original weights. If $p=0$, one obtains an equally-weighted index, where all constituents now have an index weight of $1/100$.  Thus, the smaller $p$ is chosen, the more diversified is the index; and the larger $p$ is chosen, the more capitalization-weighted is the index.

\begin{exa}
Let us do a quick example, but with two instead of 100 stocks. Assume the original index weights are $\mu_1 = 0.7$ and $\mu_2 = 0.3$.   Consider $p=1/2$.  Then the new index weights are
\begin{align*}
	\eta_1 = \frac{\sqrt{0.7}}{\sqrt{0.7} + \sqrt{0.3}} \approx 0.60; \qquad
	\eta_2 = \frac{\sqrt{0.3}}{\sqrt{0.7} + \sqrt{0.3}} \approx 0.40.
\end{align*}
If we had chosen $p=3/4$ then we would get
\begin{align*}
	\eta_1 = \frac{0.7^{3/4}}{0.7^{3/4} + 0.3^{3/4}} \approx 0.65; \qquad
	\eta_2 = \frac{0.3^{3/4}}{0.7^{3/4} + 0.3^{3/4}} \approx 0.35.
\end{align*}
\end{exa}

These power weights and similar constructions have been studied in the literature of Stochastic Portfolio Theory; see for example,  \cite{Fe} or \cite{Karatzas:Ruf:2016}. These weights also have the advantage to collect a `diversity premium' \cite{Banner:diversification}.

One might want to consider a slightly more involved transformation of the index weights in order to avoid overweighting  small index components and also to reduce the necessary trading during  rebalancing.  Thus, instead of transforming the index weights by taking their power and normalizing these numbers, one could linearize the power transformation near zero; hence, in particular, only transforming the larger weights (while making sure that monotonicity is preserved).

\bibliography{aa_bib}{}
\bibliographystyle{apalike}

\end{document}